\documentclass[manuscript]{aastex}

\usepackage{amsmath}

\shorttitle{Intermediate-degree mode frequency shifts during solar minimum}
\shortauthors{Tripathy et al.}

\begin{document}

\title{UNUSUAL TRENDS IN SOLAR P-MODE FREQUENCIES DURING THE CURRENT EXTENDED MINIMUM}
\author{S. C. Tripathy, K. Jain, F. Hill, J. W. Leibacher}
\affil{National Solar Observatory, Tucson, AZ 85719, USA}
\email{stripathy@nso.edu}

\begin{abstract}
We investigate the behavior of the  intermediate-degree  
mode frequencies of the sun during the current extended minimum phase to explore 
the time-varying conditions in the solar interior. Using contemporaneous helioseismic 
data from GONG and MDI, we find that the changes in resonant mode frequencies during the activity minimum 
period are significantly greater than the changes in solar activity as measured by 
different proxies.  We detect a seismic minimum in MDI {\it p}-mode frequency shifts 
during 2008 July--August  but no such signature is seen in mean shifts computed from GONG 
frequencies.  We also 
analyze the frequencies of individual oscillation modes from GONG data as a function 
of latitude  and observe a signature of the onset of the solar cycle 24 in early 2009. 
Thus the intermediate degree modes do not confirm the onset of the cycle 24  during late 2007
as reported from the analysis of the low-degree {\it GOLF} frequencies. Further, both the GONG and MDI frequencies show a surprising anti-correlation between frequencies and activity proxies 
during the current minimum, in contrast to the behavior during the minimum between cycles 
22 and 23. 
\end{abstract}


\section{INTRODUCTION}
The current solar minimum, between cycles 23 and 24, has been unusually long and has provided 
us with a unique opportunity to characterize the quiet sun.  
It is also the first deep minimum to be observed with  modern instrumentation and techniques
including helioseismology.  
In order to understand what causes and sustains such a prolonged period with minimal or no 
 solar activity, as commonly  measured by the number of sunspots on the visible solar disk, 
numerous efforts are underway  to find clues to such an unusual behavior. 
In this context, we use the frequencies of the global oscillation modes of the sun 
to learn about the seismic conditions of the solar interior  during the current minimum phase.

The variation of the oscillation frequencies with solar  cycle 
has been the subject of many studies and is now well established \citep[see, for example,]
[and references therein]{jain09}.  
However, the physical origin of these changes is still an active field of research. The 
extended minimum activity epoch has provided an important period to analyze and interpret the 
frequency variations during quiet periods of solar activity. \citet{brom09} have 
analyzed  the low-degree, (0 $\le$ $\ell$ $\le$ 3) mode frequencies over the last three solar cycles using  
{Birmingham Solar Oscillations Network} (BiSON) frequencies, and they find that the level of the 
present minimum is significantly deeper in the mode frequencies than in the surface activity
observations.  \citet{david09} have examined the low-degree modes from the Global Oscillations at Low Frequency ({\it GOLF}) time series.  They report that the frequency 
shifts of the $\ell$ = 0 and $\ell$ = 2 modes show an increase starting in the second half of 2007 while the $\ell$ = 1 mode follows the general decreasing trend of the solar surface activity and  
interpret the different behavior as arising from different geometrical responses to the spatial distribution of the magnetic field beneath the solar surface. Based on this, they 
argue that solar cycle 24 started during the last quarter of 2007. The result is quite intriguing  since no such evidence is seen in the BiSON data.  Figure~1 of \citet{brom09} shows that both the frequency shifts and activity as measured 
by the 10.7 cm radio flux (hereafter, $F_{10.7}$)  continue to decrease through 2009 April.
Also, the preliminary investigation of  \citet{jain10} using 36-d GONG frequencies did not detect 
the onset of the new cycle until the end of 2008. 
However, the   torsional oscillations inferred from medium-degree 
{\it p}-modes indicated that the minimum was reached during 2008 \citep{howe09}. 

Figure~1 shows the current level of solar surface activity  as measured by four 
different proxies: $F_{10.7}$, the International sunspot number\footnote{both $F_{10.7}$ and  $R_I$ are obtained from \url{http://www.ngdc.noaa.gov/STP/SOLAR/getdata.html}}, $R_I$,
the Total Solar Irradiance \citep[TSI;][]{sorce} and Magnetic Plage Strength Index \citep[MPSI;][]{ulrich91}.  We plot the mean monthly variations for the period 2007 January to 2009 September and observe that there are significant 
differences in the behavior of different activity proxies. For example, $F_{10.7}$ shows an increase in activity beyond 2008 July signaling the onset of a new cycle while no such clear trend can be observed in TSI.  
In contrast, the sunspot number shows a deep minimum in 2009 August with a monthly mean value of zero as no sunspots were observed that month.  The indices, however, display a uniform behavior when we plot the smoothed monthly mean activities (solid lines) obtained from a boxcar average of 13 points. All of the activity proxies show    
a  rising trend since the beginning of 2009 January indicating that the solar minimum has 
already occurred. A recent  analysis also confirms that the minimum in activity, as measured by the absolute solar EUV flux, occurred on 2008 November, 28 \citep{did09}.

Given the argument of \citet{david09} that the even, low-degree modes sense the onset of the new solar cycle during late 2007 and the absence of such a signature in BiSON data has motivated us to critically examine the behavior of medium-degree  mode frequencies 
with a particular emphasis on activity minimum phases. 
In this letter, we present results obtained from 
the  {\it p}-mode acoustic frequencies 
measured by the Global Oscillation Network Group (GONG) and the {Michelson Doppler Imager} (MDI) onboard {\it Solar and Heliospheric Observatory} ({\it SOHO}) spacecraft.  We also investigate the latitudinal variations of the shifts obtained from the frequencies of individual $n, \ell, m$ modes, where $n$ is the radial order and $m$ is the azimuthal order.      
 
\section{ANALYSIS AND RESULTS}
The MDI mode frequencies\footnote{\url{http://quake.stanford.edu/$\sim$schou/anavw72z/}} that we use consist of 66 sets, each of which is calculated from a
time series of 72 days \citep{schou99} and covers  the period between 1996 May 1 and 2009 September 26.  We also use 72-day long GONG time series to calculate the centroid frequency, $\nu$, for each ($n, \ell$) multiplet \citep{hill96} for each of 71 non-overlapping sets that covers the period between  1995 May 7 and  2009 May 4, a few months shorter than the MDI frequencies.
The mean frequency shift, $\delta\nu$, for each data set is obtained as a error-weighted mean 
where the error corresponds to the formal fitting uncertainties returned 
by the fitting procedure. The frequencies were also weighted by the mode inertia before 
the mean was computed \citep[For details see][]{sct07}. 
We analyze only those modes that are present in both the data sets. 
The included modes cover a frequency range of 1500~$\mu$Hz $\le$ $\nu$ $\le$ 3600~$\mu$Hz and degree range of  
22 $\le$ $\ell$ $\le$ 150 and sample the outer 30\% of the solar interior. 
 For calculating the frequency shifts, we construct a reference frequency  
 as an average over the frequencies of a given multiplet present in all MDI data sets. 
The same set of reference frequencies are used to calculate the mean shifts from the GONG 
data. This choice introduces a constant offset between MDI and GONG shifts and does not affect 
the results.
 
As a proxy for the solar activity, we choose $F_{10.7}$  
which represents a combination of sunspots, 
radio plages and quite-sun background emission \citep{kundu65}  and the 
International sunspot number, $R_I$. The former is chosen so that our results can be compared 
with those of \citet{brom09} and \citet{david09}, while the later is chosen because it acts  
as a universal activity indicator. Figure~2 shows the temporal variations of the GONG and MDI frequency shifts  
and the linearly scaled $F_{10.7}$ averaged over the same time period 
as the length of the time series. On the time scale 
of the solar cycle, frequency shifts and solar activity exhibit the well established 
linear relationship as measured by the Pearson's 
correlation coefficient,  $r_p$ = 0.99 for both GONG and MDI.

 For a comprehensive and comparative analysis between previous and current activity minima, 
the GONG frequency shifts and the corresponding $F_{10.7}$ during these two periods are 
shown in Figure~3.  
The dash-dot and dash-dot-dot-dot lines in  Figure~3b indicate the   
minimum value in activity and frequency shifts corresponding to the previous cycle, respectively
and indicate differences between the current and last minima. Both $F_{10.7}$ 
and frequency shifts appear weaker this minimum  than  the previous; the decrease in 
mean frequencies and activity are of the order of 11\% and 4\%, respectively 
implying a  significant  change 
in oscillation frequencies as compared to the solar activity. A similar result seen 
in BiSON low-degree modes led \citet{brom09} to conclude that the changes in frequencies 
are deeper than the surface activity.  During the extended minimum we also observe discrepancies in the behavior of $\delta\nu$ and $F_{10.7}$. The oscillation frequencies 
continue to decrease through 2009 April (end of data)  while the solar activity as measured by $F_{10.7}$ 
shows a minimum near the end of 2008 and an increasing trend thereafter. This reflects 
 a deviation from the assumed linear relationship between the two indicating 
an anti-correlation.  No such behavior is seen in the 
previous minimum period (Figure~3a). Therefore  the relationship between $\delta \nu$ and solar activity proxies during the current extended minimum phase appears to be  more complex than the previous minimum between cycles 22 and 23.   

 Since different activity proxies behave differently during the 
activity cycle (Figure~1) and some proxies are better correlated with frequency shifts 
than others \citep{chaplin07, jain09}, we compare 
the variation of MDI frequency shifts with two different proxies, $F_{10.7}$ (Figure~4a) and $R_I$ (Figure~4b), for the present minimum period. 
  It is seen that the frequency shifts are in better agreement with $R_I$ rather than $F_{10}$ with a marginally higher correlation 
coefficient. This does not agree with the argument 
that the frequency shifts are more sensitive to both the strong and weak components of the 
magnetic field \citep{chaplin07, jain09}. Thus, we see additional evidence 
of the complex relationship between frequency shifts and activity proxies in different phases of the solar activity and particularly during the extended minimum period.
 
 Furthermore, we note an anti-correlation between the MDI frequency shifts and both of the 
activity indices starting from the end of 2007 ($r_p$~$\approx$~$-$~0.1) 
similar to those seen in  GONG frequency shifts.  
The correlation coefficient decreases to  $-$~0.5 if we consider the variation between 2008 July and the end of the data.  Although the value of the coefficients are not significant, the 
signature that the frequency shifts and solar activity proxies are  opposite in phase is  important since this has not been seen in previous cycles. Therefore, the extended minimum period appears to be rather unusual since all earlier studies involving the frequency shifts report close correlations between the frequency shifts and solar surface activity proxies.   

Figure~4 reveals a minimum in frequencies during the 2008 May--July period and a minimum in $R_I$
during 2008 July-August which is earlier by a few months in comparison to the minimum seen in $F_{10.7}$. Both of the activity proxies suggest that the solar minimum has already occurred 
with a phase lag of nearly two  and five months between $\delta\nu$, and $R_I$ and $F_{10}$, 
respectively. However, we note that $F_{10.7}$ is showing a downward trend during 2009 September 
and if this downward trend continues, the  minimum in activity  may shift to a later period.  

Since the oscillations are sensitive to the conditions beneath the solar surface, it is not surprising that the signature of the onset of solar cycle 24  is first visible in the oscillation data. The absence of such a signature in GONG data could be related to the mode fitting techniques used by the data reduction pipelines as has been demonstrated  
in earlier studies \citep{schou02, sb03}.  The GONG technique fits  one individual mode 
($n$, $\ell$, $m$) at a time together with a number of leaks from other $\ell$ but ignores 
known leaks from modes with the same ($n$, $\ell$) but different $m$.  The MDI algorithm, on the 
other hand,  fits all modes of all $m$ for a given multiplet ($n$, $\ell$) simultaneously 
and uses a combination of leakage matrix and the known parameters from the fitting 
of other modes to estimate the leaks from these modes.

The GONG mode-fitting algorithm, in addition to the rotation corrected frequencies, also produces 
frequencies of individual modes as function of $n$, $\ell$, $m$, where $m$ represents the number of nodal lines around the equator \citep{hill96}.  It is possible to follow the changes in
oscillation modes as a function of the latitudes using different values of $m$/$\ell$. The sectoral ($|m|/\ell$ = 1) and near sectoral modes are sensitive to  the region near the 
equator while zonal ($|m|/\ell$ = 0) and tesseral modes (0 $<$ $|m|/\ell$ $<$ 1) sense a wider 
range of latitudes away from  the equator \citep{komm02}.   
In Figure~5 we show the mean relative frequency shifts, $\delta\nu_{nlm}$, in the 5-min band (2800 $\le$ $\nu$ $\le$ 3200; 20  $\le$ $\ell$  $\le$ 100) 
for four  different $|m|/\ell$ values (0.5 $\le |m|/\ell \le$ 0.8) representing mid latitudes 
 during the extended minimum period; the mean shifts are calculated with reference to the model frequencies used by the GONG pipeline.  It is evident that the frequency shifts during the current minimum phase  are greater as compared to the  minimum in 1996 (shown by the dotted lines in each panel) and continue a downward trend for $|m|/\ell$ = 0.5 and 0.8.  In contrast,  for the other two values of $|m|/\ell$  that are closer to the latitudinal belt where solar activity first appears during the onset of a solar cycle, 
the shift indicates an increasing trend starting from early 2009. This may be the first seismic signature of the beginning of the solar cycle 24 since this epoch is closer to the period of minimum solar activity.  The evidence, however, is based on one data point and will  require additional data for  confirmation.

\section{SUMMARY}
Using contemporaneous helioseismic 
frequencies from GONG and MDI, we investigated the behavior of the  
oscillation frequencies of the sun during the extended minimum phase of the current solar cycle.  
We find that the  state of the solar interior as measured by the 
oscillation frequencies is  varying significantly in time even if the traditional measures of 
solar surface activity, such as sunspot number, magnetic field strength indices, or the 10.7~cm radio 
flux are nearly constant.  The MDI frequency shifts indicate a ``seismic" minimum around mid 2008 
while no such signature is seen in the GONG data. However, the relative frequency shifts calculated from  individual GONG multiplets hints at the start of the solar cycle 24 in early 2009.

Although low degree modes carry information from the deep interior, their dwell time near the surface 
is higher than in the core because the sound speed inside the sun increases rapidly with depth. 
Thus, both low and intermediate degree modes are most influenced by the conditions near  
the solar surface. Therefore it is a puzzle that the low degree mode
frequencies measured from the {\it GOLF} data support an early onset of solar cycle 24 
during the last quarter of 2007 \citep{david09} while no such evidence is seen  in the 
BiSON low-degree modes \citep{brom09} or the intermediate-degree modes from GONG.  MDI 
frequencies, on the other hand, indicate a seismic minimum but the period is about 
one and half year later than seen in the  
{\it GOLF} data.  It is important to note that the frequency shifts calculated from intermediate degree modes or BiSON low degree modes represent averages over all modes whereas the shifts obtained from {\it GOLF} data correspond to individual low degree modes.
Moreover, both the GONG and MDI frequencies show a surprising anti-correlation between frequencies and activity proxies 
during the current minimum, in contrast to the behavior during the previous minimum between 
cycles 22 and 23 indicating that the current minimum is unusual. This also suggests that in addition to the surface 
magnetic  
activity, the variations in oscillation frequencies may be caused by some other effects e.g. changes in the 
structure of the Sun below the photosphere.

\acknowledgments
This work utilizes data obtained by the Global Oscillation Network
Group program, managed by the National Solar Observatory, which
is operated by AURA, Inc. under a cooperative agreement with the
National Science Foundation. The data were acquired by instruments
operated by the Big Bear Solar Observatory, High Altitude Observatory,
Learmonth Solar Observatory, Udaipur Solar Observatory, Instituto de
Astrof\'{\i}sica de Canarias, and Cerro Rollo Interamerican
Observatory. The work also utilizes data from MDI onboard {\it SOHO}. 
{\it SOHO} is a  project of international cooperation between ESA and NASA. 
The study also includes data from the synoptic program at the 150-Foot Solar Tower of the Mt. Wilson
Observatory. The Mt. Wilson 15-Foot Solar Tower is operated by UCLA, with funding from NASA, ONR, and 
NSF, under agreement with the Mt. Wilson Institute. 
The work is supported  by NASA grant NNG 08EI54I.

{}
\onecolumn
\begin{figure}
\includegraphics[scale=0.68,angle=90]{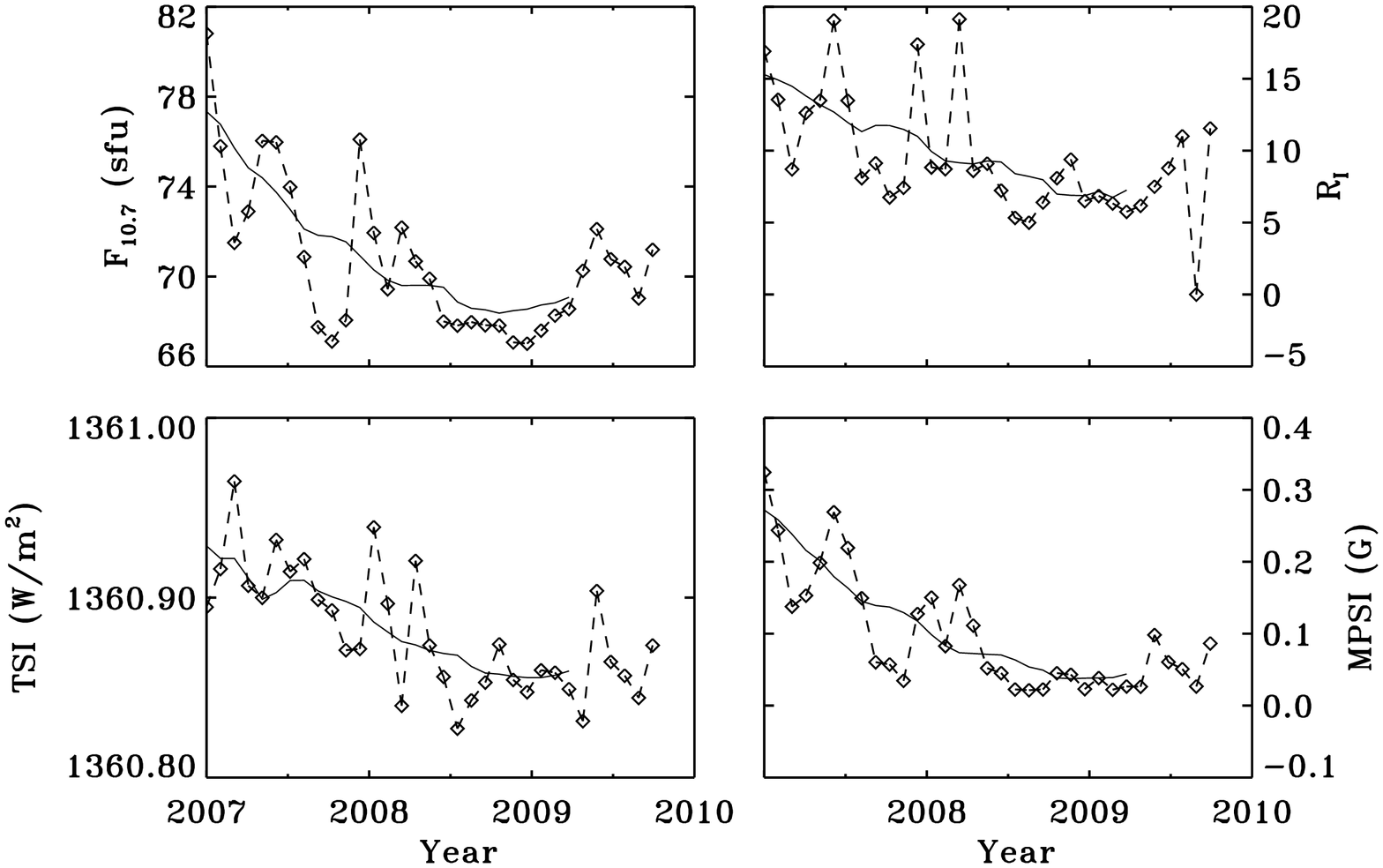}
 \caption{Average monthly numbers of different activity proxies (symbols) for the period 
2007 January until 2009 September. The solid line in each panel represents the 
smoothed monthly mean values obtained from a boxcar average of 13 points.   
\label{fig1}}
\end{figure}

\begin{figure}
\plotone{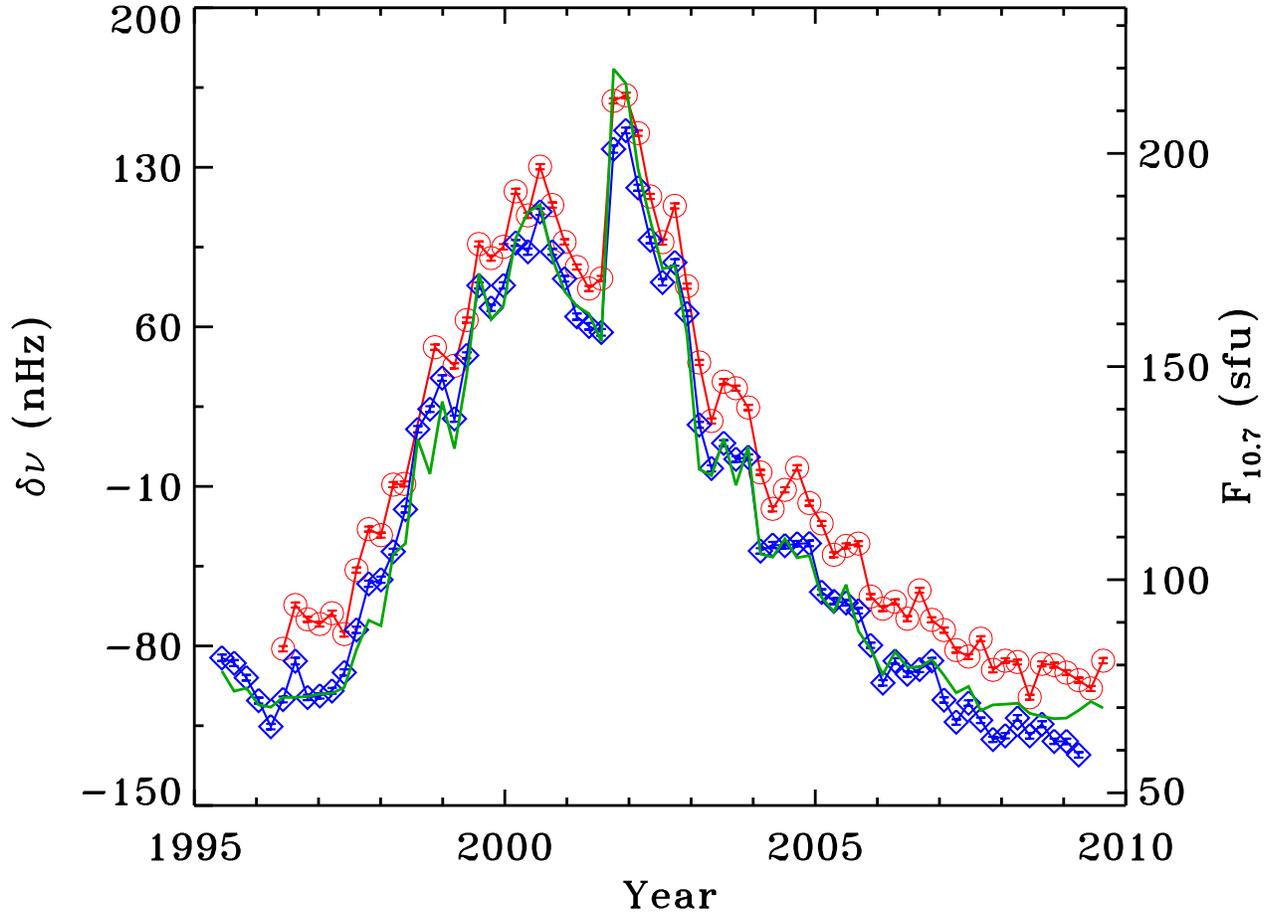} 
 \caption{Temporal evolution of the intermediate degree frequency shifts from GONG (diamonds) and MDI  (circles). The symbols have been joined by a line to visualize the solar cycle.  The solid 
line represents the 10.7 cm radio flux scaled linearly with the GONG frequency shifts.  
\label{fig2}}
\end{figure}

\begin{figure}
\plottwo{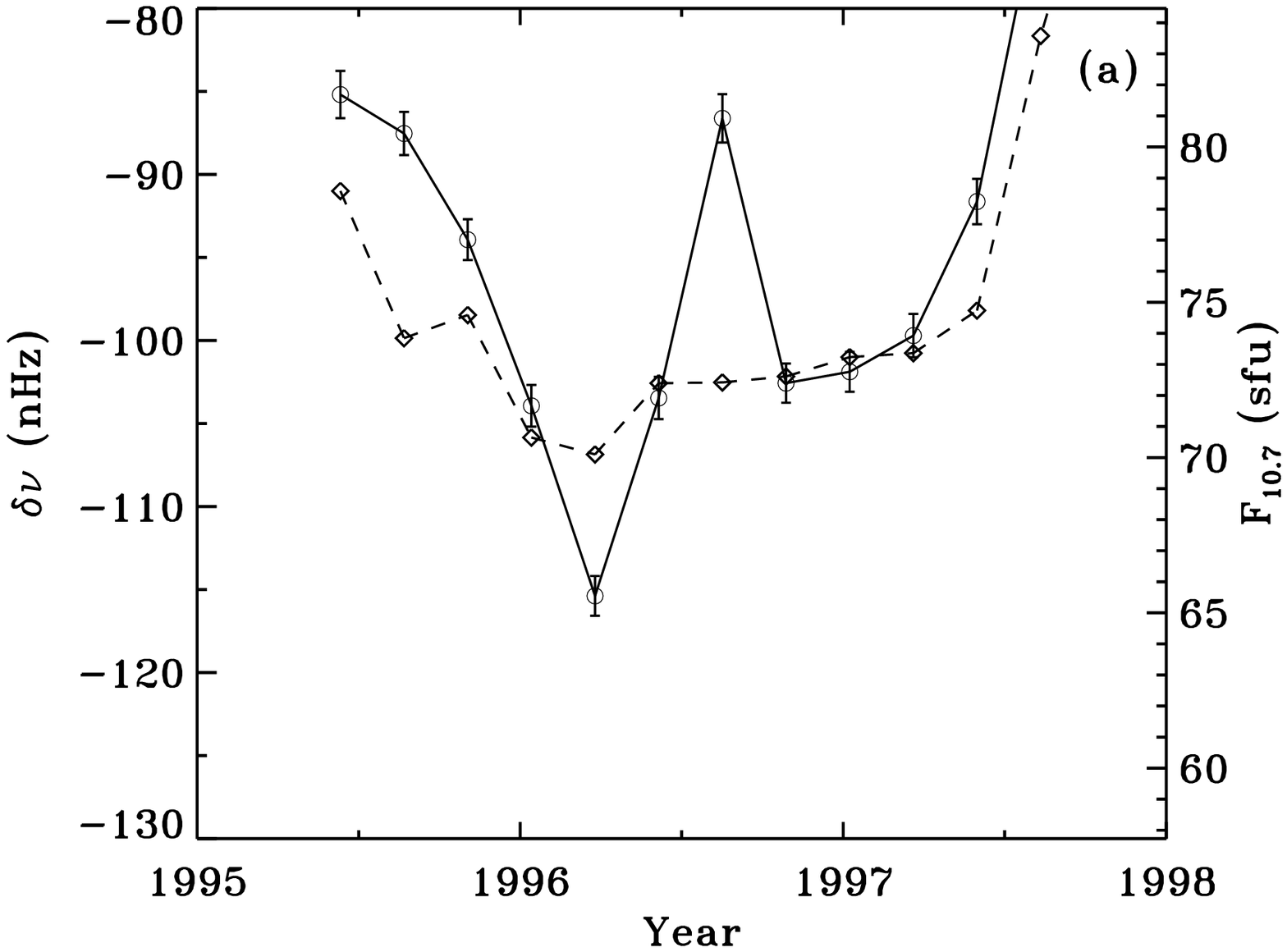}{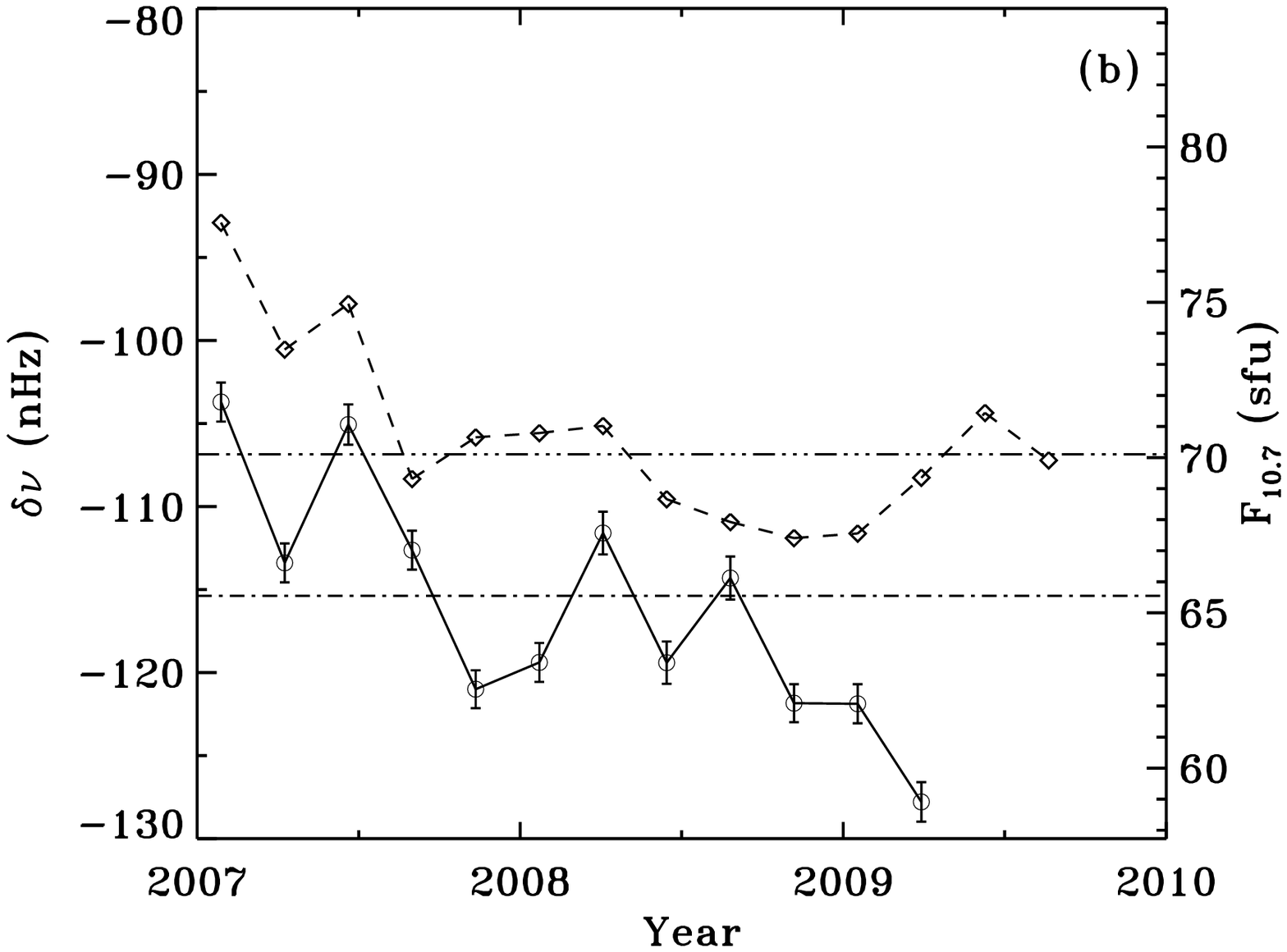} 
\caption{Temporal evolution of GONG frequency shifts (circles)  
and $F_{10.7}$ (diamonds) during the (a) previous and (b) present minimum phase of the activity cycle.   The dash-dot and the dash-dot-dot-dot lines in panel (b) indicate $F_{10.7}$ and  the frequency shifts corresponding to the minimum phase between cycles 22 and 23, respectively.
\label{fig3}}
\end{figure}

\begin{figure}
\plottwo{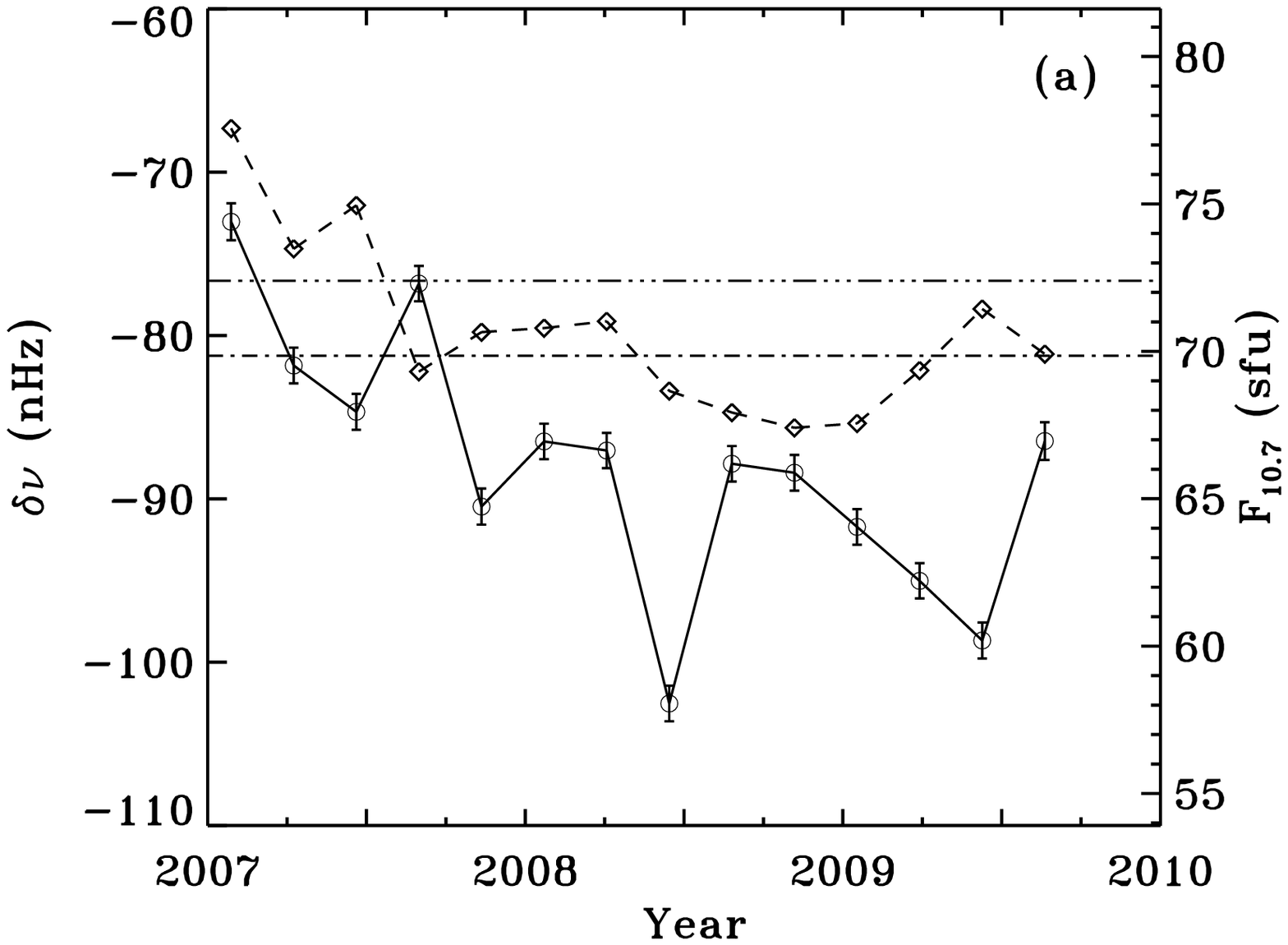}{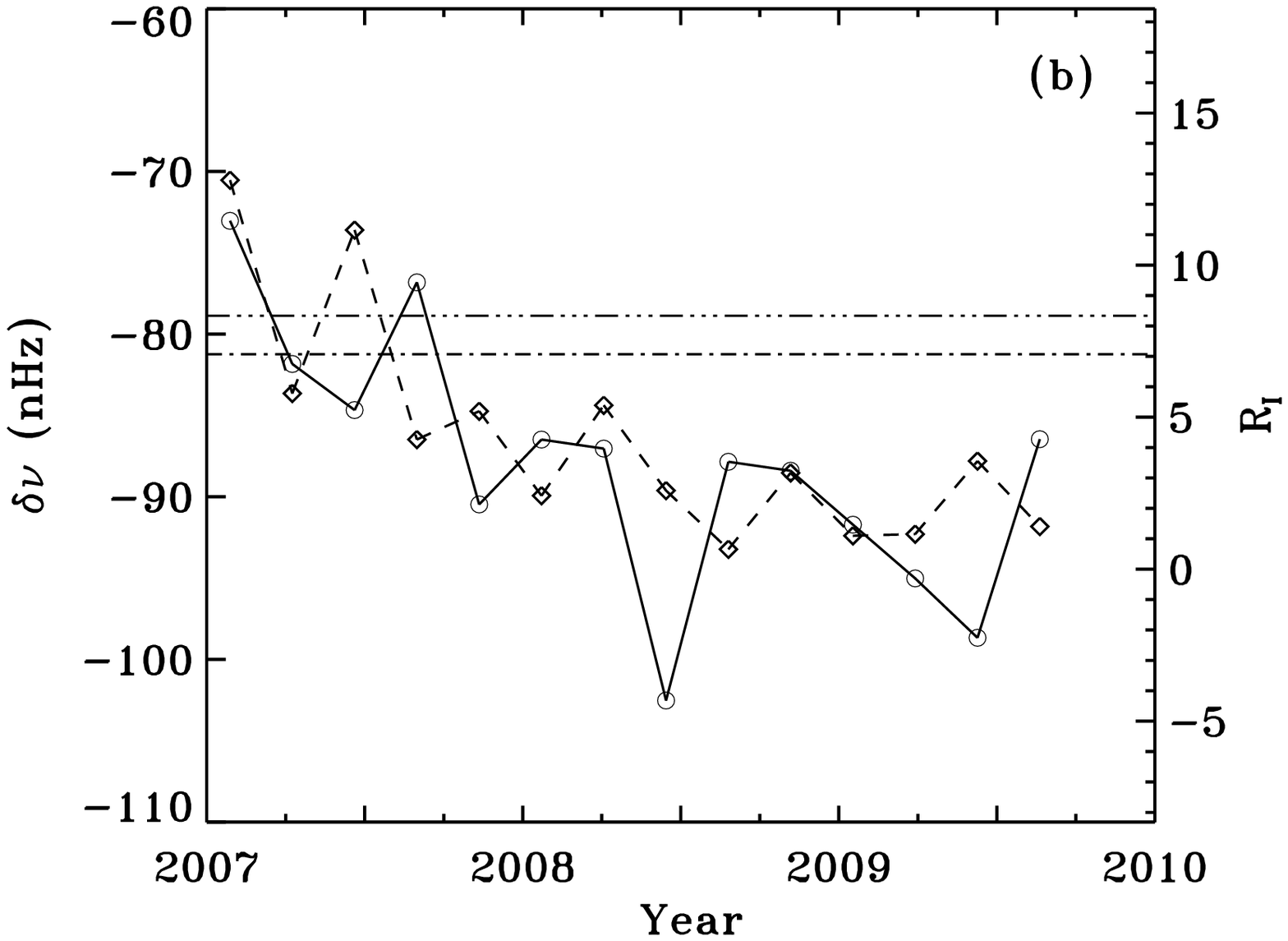}
 \caption{The temporal evolution of MDI frequency shifts during the current minimum phase of the solar cycle for two different activity proxies, (a) $F_{10.7}$ and (b) $R_I$. 
The dash-dot and the dash-dot-dot-dot lines in each panel indicate the solar activity and the frequency shifts corresponding to the minimum phase between cycles 22 and 23, respectively.
The Pearson's correlation coeffcient between 
$\delta\nu$, and $F_{10.7}$ and $R_I$ are 0.4 and 0.5, respectively. 
 \label{fig4}}
\end{figure}

\begin{figure}
\begin{center}
\plotone{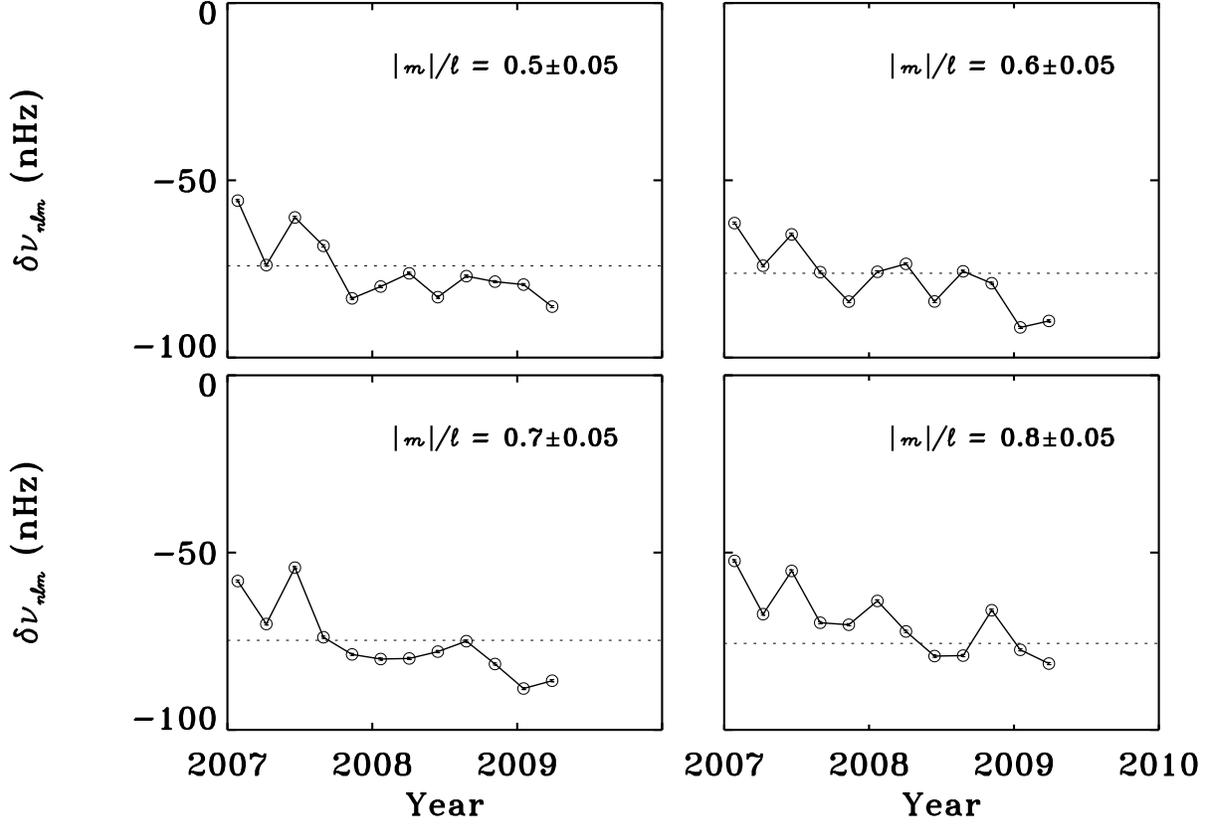}
 \caption{Temporal evolution of mean frequency shifts at selected $|m|/\ell$ values.  The frequencies were weighted by the inverse mode inertia 
before being averaged. The dotted line in each panel denote the minimum frequency shifts for the same $|m|/\ell$ during 1996.     
\label{fig5}}
\end{center}
\end{figure}

\end{document}